\documentclass[iop]{emulateapj}
                                                                                                                     
\usepackage{epsfig}
\usepackage{hyperref}
\usepackage{natbib}
\usepackage{graphicx}
\usepackage{amsmath}
\usepackage{multirow}
\usepackage{float}
\usepackage[none]{hyphenat}
\usepackage[export]{adjustbox}

\def\simgt{\lower.5ex\hbox{$\; \buildrel > \over \sim \;$}}
\def\simlt{\lower.5ex\hbox{$\; \buildrel < \over \sim \;$}}

\def\amin{\ifmmode^{\prime}\else$^{\prime}$\fi}
\def\asec{\ifmmode^{\prime\prime}\else$^{\prime\prime}$\fi}

\def\simgt{\lower.5ex\hbox{$\; \buildrel > \over \sim \;$}}
\def\simlt{\lower.5ex\hbox{$\; \buildrel < \over \sim \;$}}

\newcommand\chandra{{\it Chandra}}

\newcommand\xmm{{\it XMM-Newton}}

\newcommand\nustar{{\it NuSTAR\/}}
\newcommand\fermi{{\it Fermi\/}}

\shorttitle{NuSTAR Observation of IC 443}
\shortauthors{S. Zhang et al.}

\begin{document}

\title{{\it NuSTAR} Detection of a Hard X-ray Source in the \\Supernova Remnant - Molecular Cloud Interaction Site of IC~443}

\author{Shuo Zhang\altaffilmark{1} Xiaping Tang\altaffilmark{2}, Xiao Zhang\altaffilmark{3}, Lei Sun\altaffilmark{3}, Eric V. Gotthelf\altaffilmark{4}, Zhi-Yu Zhang\altaffilmark{5,6}, Hui Li\altaffilmark{1}, Allen Cheng\altaffilmark{7}, Dheeraj Pasham\altaffilmark{1}, Frederick K. Baganoff\altaffilmark{1}, Kerstin Perez\altaffilmark{7}, Charles J. Hailey\altaffilmark{4}, Kaya Mori\altaffilmark{4}}

\altaffiltext{1}{MIT Kavli Institute for Astrophysics and Space Research, Cambridge, MA 02139, USA; shuo@mit.edu}
\altaffiltext{2}{Max Planck Institute for Astrophysics, Karl-Schwarzschild-Str. 1, D-85741 Garching, Germany}
\altaffiltext{3}{School of Astronomy and Space Science, Nanjing University, Nanjing 210023, China}
\altaffiltext{4}{Columbia Astrophysics Laboratory, Columbia University, New York, NY 10027, USA}
\altaffiltext{5}{Institute for Astronomy, University of Edinburgh, Royal Observatory, Blackford Hill, Edinburgh EH9 3HJ, UK}
\altaffiltext{6}{ESO, Karl-Schwarzschild-Str.~2, D-85748 Garching, Germany}
\altaffiltext{7}{Department of Physics, Massachusetts Institute of Technology, Cambridge, MA 02139, USA}

\begin{abstract}
 
We report on a broadband study of a complex X-ray source (1SAX J0618.0+2227) associated with the interaction site of the supernova remnant (SNR) IC~443 and ambient molecular cloud (MC) using \nustar, \xmm, and \chandra\ observations. 
Its X-ray spectrum is composed of both thermal and non-thermal components.
The thermal component can be equally well represented by either a thin plasma model with $kT=0.19$~keV or a blackbody model with $kT=0.11$~keV.
The non-thermal component can be fit with either a power-law with $\Gamma \sim1.7$ or a cutoff power-law with $\Gamma \sim 1.5$ and a cutoff energy at $E_{cut} \sim 18$~keV.
Using the newly obtained \nustar\ dataset, we test three possible scenarios for isolated X-ray sources in the SNR-MC interaction site: 1) pulsar wind nebula (PWN); 2) SNR ejecta fragment; 3) shocked molecular clump.
We conclude that this source is most likely composed of a SNR ejecta (or a PWN) and surrounding shocked molecular clumps. 
The nature of this hard X-ray source in the SNR-MC interaction site of IC 443 may shed light on unidentified X-ray sources with hard X-ray spectra in rich environments for star forming regions, such as the Galactic center.
\end{abstract}
\keywords{X-rays: individual (IC 443) --- X-rays: ISM --- ISM: supernova remnants}

\section{Introduction}

Core-collapsed supernovae are expected to interact with the parent clouds which form the progenitor star. 
There are dozens of established supernovae remnant and molecular cloud (SNR-MC) interaction systems, which were identified with observational signatures like OH 1720 MHz maser emission and molecular line broadening \citep{Jiang2010}. 
$\gamma$-ray emission has also been detected from some of the SNR-MC interaction systems, pointing to hadronic processes \citep{Acero2016}. 
Among them, IC~443 (G~189.1+3.0), an evolved SNR with a diameter of 45\amin\ and a distance of 1.5~kpc, serves as an ideal laboratory to study SNR-MC interaction in details because of its brightness and proximity (e.g. \citealp{Fesen1980}). 
IC~443 interacts with surrounding neutral interstellar medium of complex structures,
which has been confirmed by the shock-broadened emission lines of H{\sc i}, CO, H$_2$O, etc. (e.g. \citealp{Burton1988, VD1993, Cesarsky1999}) and the detection of OH 1720 MHz masers and CO clumps (e.g. \citealp{Claussen1997, Hoffman2003, Hewitt2006}). 

IC~443 radiates broadband emission from radio to $\gamma$-rays.
The X-ray emission from IC~443 is composed of extended thermal X-ray emission components and a number of isolated hard X-ray sources (e.g. \citealp{Petre1988, Wang1992, AA1994, Keohane1997, BB2000, Kawasaki2002, BB2003, Bykov2005, Troja2006, Troja2008, Bocchino2008}).
{\it BeppoSax} detected two bright isolated X-ray sources, 1SAX J0617.1+2221 and 1SAX J0618.0+2227, from the IC 443 region \citep{BB2000}.
1SAX J0617.1+2221 is located to the south of the SNR and has been identified as a pulsar wind nebula (PWN)  \citep{Olbert2001, BB2001}.
1SAX J0618.0+2227 resides in the SNR-MC interaction site tracing the southeast boundary of the SNR.
\xmm\ resolved 1SAX J0618.0+2227 into three localized X-ray sources (XMMU J061804.3+222732, XMMU J061806.4+222832, XMMU J061759.3+222739, dubbed as Src 1-3) with hard X-ray spectra and surrounding diffuse emission \cite{Bykov2008}.

Several mechanisms have been proposed to explain X-ray emission from the SNR-MC interaction systems.
Firstly, broadband non-thermal diffuse emission from molecular clouds in an SNR-MC system could be attributed to synchrotron and bremsstrahlung emission from energetic leptons, either primary leptons directly accelerated by SNRs or secondary leptons produced by proton-proton ($p$-$p$) interaction \citep{Bykov2000, Vink2008, Gabici2009, Tang2011}. 
For isolated X-ray sources in the SNR-MC interaction sites, they could bear the nature of 1) fast-moving metal-rich SNR ejecta fragments colliding into molecular clouds \citep{Bykov2002, BB2012}, 2) shocked molecular clumps \citep{BB2000} or 3) PWNe.

In the $\gamma$-ray band, a GeV source 3EG J0617+2238 ($\rm R.A.=6^{h}17^{m}18^{s}, Decl.= 22^{\circ}35\amin08\asec$ (J2000)) has been detected by {\it Fermi} from IC~443, centering on a position close to the SNR-MC interaction region \citep{Abdo2010}. 
Whether 1SAX J0618.0+2227 is the X-ray counterpart of the GeV source is another open question.
Hard X-ray observation plays an important role bridging the soft X-ray band and the GeV band.
{\it INTEGRAL}/JEM-X revealed hard X-ray emission from 1SAX J0618.0+2227 up to 35~keV, but was not able to spatially resolve it.
Therefore, aiming at resolving this hard X-ray source in the SNR-MC interaction site of IC~443, we observed 1SAX J0618.0+2227 and its ambient environment using \nustar\ during GO cycle 1.

In this paper, we report the detection and analysis of 1SAX J0618.0+2227 using the newly obtained \nustar\ data and archival \xmm\ and \chandra\ data.
The paper is organized as below.
We present the observation and data reduction methods in Section 2, the X-ray morphology of the detected source in Section 3 and its broadband X-ray spectrum in Section 4.
Lastly we discuss the emission mechanism and possible origins of this source in Section 5.

\section{Observation and Data Reduction}

1SAX J0618.0+2227, located in the SNR-MC interaction region of IC~443 was observed by \nustar\ during GO cycle-1 on 2015 December 15, with a total exposure time of 97.4~ks (obsID: 30101058002).
The region was imaged with the two co-aligned X-ray telescopes, with corresponding focal plane modules FPMA and FPMB, providing an angular resolution of $58\asec$ Half Power Diameter (HPD) and $18\asec$ Full Width Half Maximum (FWHM) over the 3--79~keV X-ray band, with a characteristic spectral resolution of 400~eV (FWHM) at 10~keV. 
The nominal reconstructed \nustar\ astrometry is accurate to $8\asec$ at 90\% confidence level.
The data were reduced and analyzed using the HEASOFT v.6.19, and filtered for periods of high instrumental background due to South Atlantic Anomaly (SAA) passages and known bad/noisy detector pixels. 
We used both FPMA and FPMB for source spectral extraction.
The resultant spectrum was grouped such that the detection significance in each data bin is at least $3\sigma$.
The confidence levels for all the error bars reported in this paper are 90\%.

To achieve a broadband joint spectral fitting for the detected source, we reanalyzed the archival \xmm\ observation of IC~443 obtained on 2006 March 30 UTC (obsID: 0301960101) with an exposure time of 81.8~ks.
We extracted source spectra from all available EPIC instruments using the \xmm\ Science Analysis Software (SAS) v.12.0.1. 
For each observation, calibrated event files were produced with the tasks \textit{epchain} and \textit{emchain} and filtered with \textit{pn-filter} and \textit{mos-filter} in order to exclude the time intervals affected by soft proton contamination. 
The source and background spectra were then extracted with the SAS \textit{mos-spectra} and \textit{pn-spectra} scripts and rebinned to have at least 30 counts in each bin to apply chi-square statistics.

In order to compare with high resolution X-ray image and spectrum, we also analyzed the archival \chandra/ASIC-S observation of the 1SAX J0618.0+2227 region taken on 2004 April 12 UTC (obsID: 4675) with an exposure time of 58.45~ks.
The source spectrum is extracted from ACIS-S using CIAO v.4.9 tool \textit{specextract}.

\section{X-ray Morphology}

\xmm\ resolved~1SAX J0618.0+2227 into three localized sources: XMMU J061804.3+222732 (Src 1), XMMU J061806.4+222832 (Src 2), and XMMU J061759.3+222739 (Src 3), as named in \cite{Bykov2008}.
For simplicity, from here below we followed the same naming to refer to the three {\it XMM} sources as Src 1-3. 
Figure 1 shows the 14\amin$\times$11\amin\ \nustar\ image in 3--40~keV of 1SAX J0618.0+2227 and its surroundings.
\nustar\ detected a bright X-ray source up to $\sim40$~keV at the location of Src 1, 
while Src 2 and Src 3 resulted in non-detection (1-$\sigma$ and 2-$\sigma$ detection respectively).
Two point sources were also detected in the field of view (FoV) of \nustar, highlighted with $r=10\asec$ white circles and noted as Ps 1 and Ps 2 here. 
The 3-40~keV image is overlaid with the 1.4~GHz continuum emission contours (magenta) \citep{Lee08, ZYZ2010}, which demonstrates the radio emission along the SNR shell.
Src 1 is $\sim0.9'$ away from a local radio emission maximum.
We also overlaid the \nustar\ image with contours (dashed yellow) of the $^{12}$CO 1--0 emission integrated over the whole velocity range, which indicates the spatial distribution of shocked H$_{2}$ gas \citep{ZYZ2010}.
Src 1-3 are located around ``clump D" \citep{VD1993}, an extended molecular clump interacting with IC~443 shell.

\begin{figure}[t]
\centering
\label{fig:image}
\includegraphics[width=0.96\linewidth]{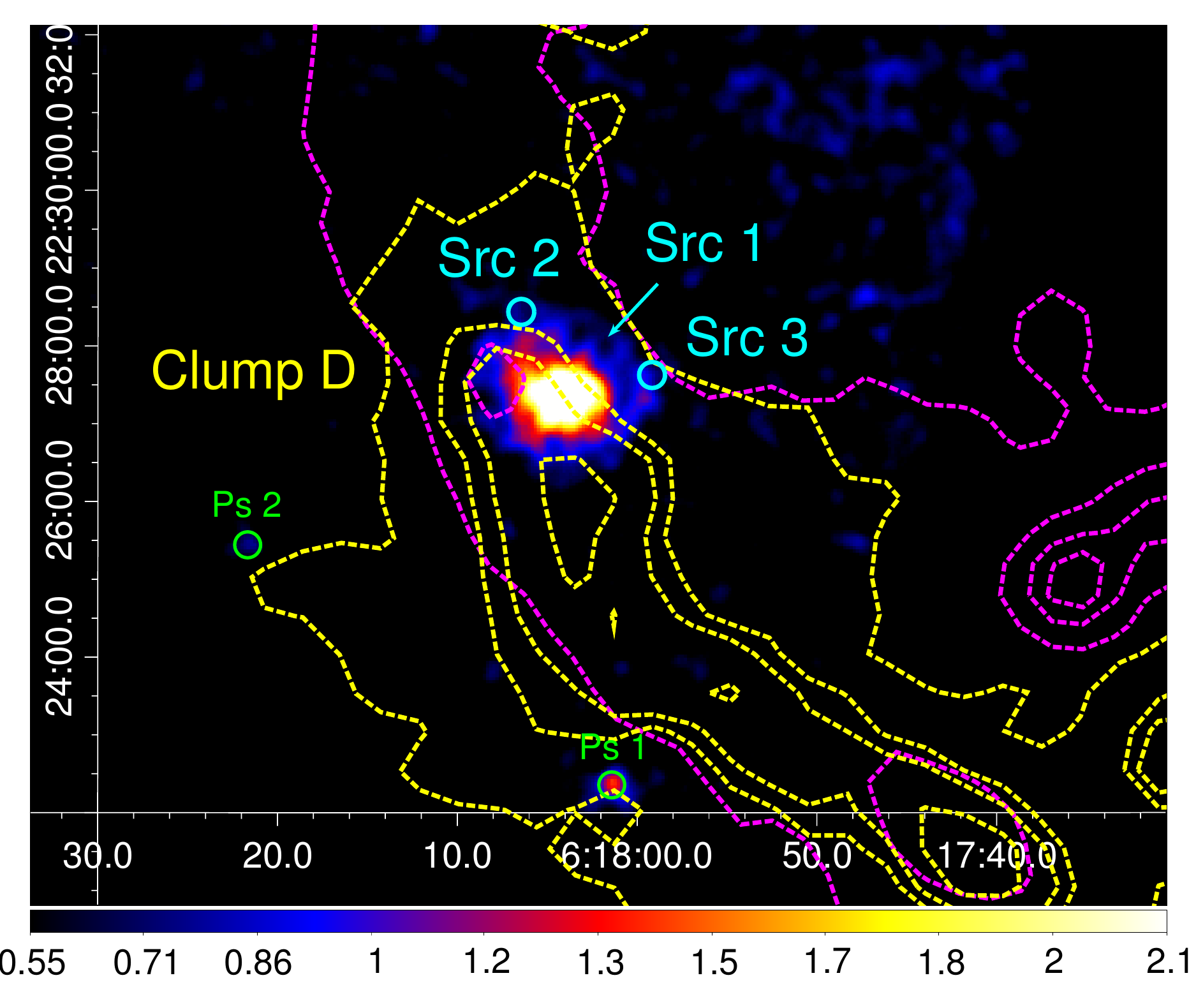}
\caption{The 13\amin\ $\times$ 18 \amin\ image of the IC~443 SNR-cloud interaction in 3--40~keV. Src 1 is the brightest source detected in the field of view; the location of undetected Src 2 \& Src 3 are outlined with $r=10\asec$ cyan circles. The image is overlaid with 1.4~GHz continuum emission contours in magenta demonstrating radio emission from SNR shell, and with dashed yellow contours of the $^{12}$CO $J$=1$\rightarrow$0 emission integrated over the whole velocity range indicating of the presence of shocked gas \citep{Lee08, ZYZ2010}. Src 1 is $\sim0.9'$ away from a local radio emission maximum, and close to clump D as defined in \citet{VD1993}.}
\end{figure}

\begin{figure}
\label{fig:chandra}
\hspace{3mm}
\includegraphics[width=1.05\linewidth]{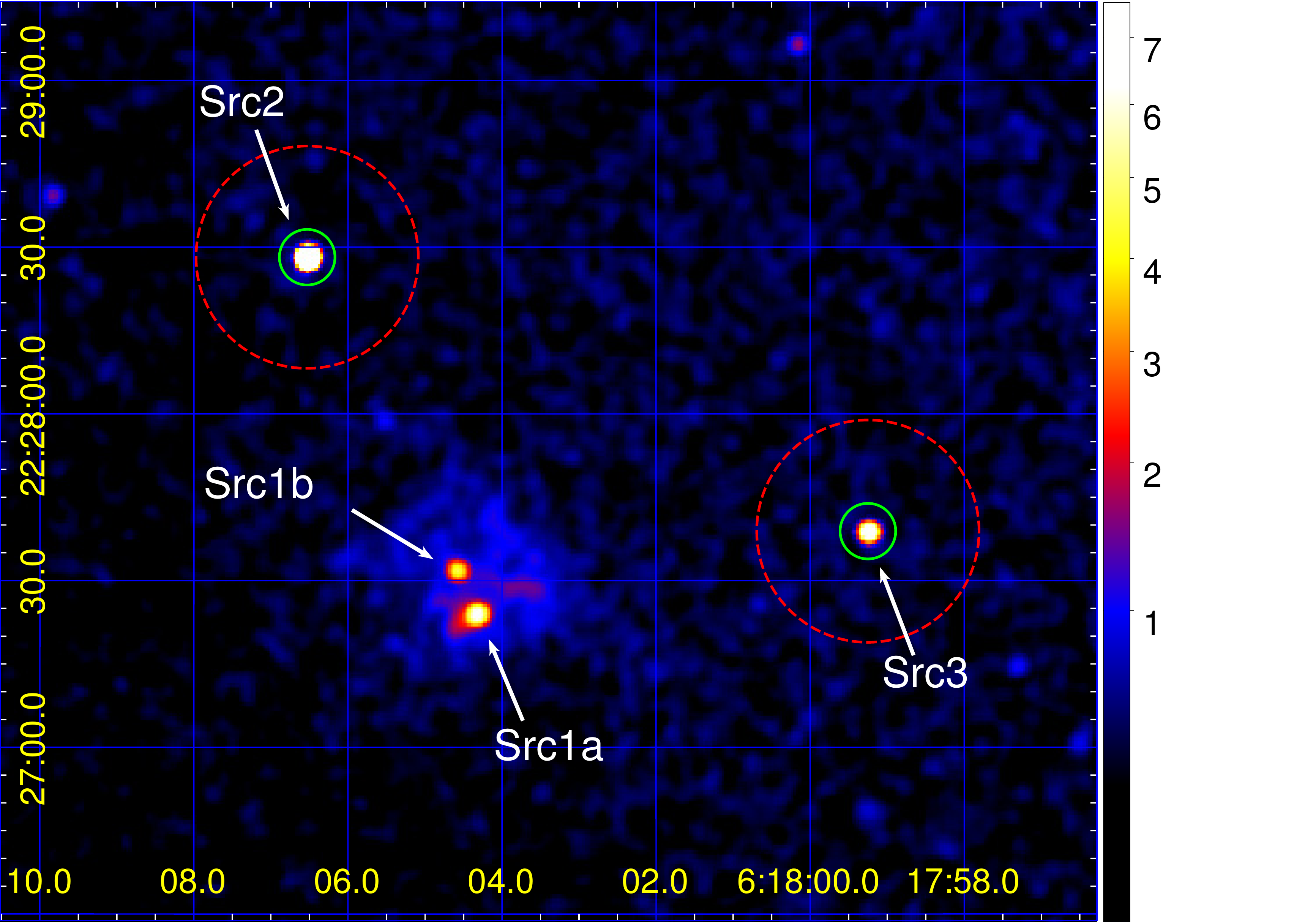}
\caption{0.3-8.0 keV image of the Src 1 region obtained by \chandra/ACIS-S on 2004 April 12 UTC, demonstrating complicated sub-structures of Src 1. Src 1 is resolved into a slightly extended source Src 1a and a point source Src 1b, both embedded in nonuniform extended emission with $\sim30\asec$ in size. Src 2 and Src 3 are outlined with green circles with $r=10\asec$, from which we extracted source spectra. The red dashed circles outlines the background regions for Src 2 and Src 3, respectively. }
\end{figure}

Figure 2 shows the zoomed-in \chandra/ACIS-S image of the vicinity of Src 1-3 in 0.3--8~keV. 
The sub-arcsecond spatial resolution of \chandra\ allows to reveal complicated substructures of Src 1.
Src 1 is resolved into a slightly extended source Src 1a and a point source Src 1b, 10\asec\ apart, both embedded in non-uniform extended emission with $\sim30$\asec\ in size. 
Src 2 and Src 3, both about 1.5\amin\ from Src 1, are outlined with green circles with $r=10$\asec. 

Figure 3 shows the zoomed-in \nustar\ images of the vicinity of Src 1 in 3--10~keV (left panel) and 10--40~keV (right panel).
Src 1 is outlined with a cyan regions with $r=30\asec$; Src 2 and Src 3 are outlined with cyan circles with $r=10\asec$.
The position of Src 1a and 1b as resolved by \chandra\ are represented by two small black circles.
It is clear that the hard X-ray source detected above 10 keV by \nustar\ is the hard X-ray counterpart of Src 1.
\nustar\ cannot resolve the complicated sub-structures of Src 1 (a few sub-clumps like Src 1a and Src 1b, embedded in more extended diffuse emission).
Considering the position uncertainty of \nustar, the centroid of the hard X-ray source is consistent with both Src 1a and Src 1b at 90\% level, though closer to Src 1a. 
Non-uniform diffuse emission in between Src 1-3 extends up to 80\asec\ from Src 1, as outlined by the magenta region.

\begin{figure*}
\centering
\label{fig:zoomin}
\includegraphics[angle=0, width=0.9\linewidth]{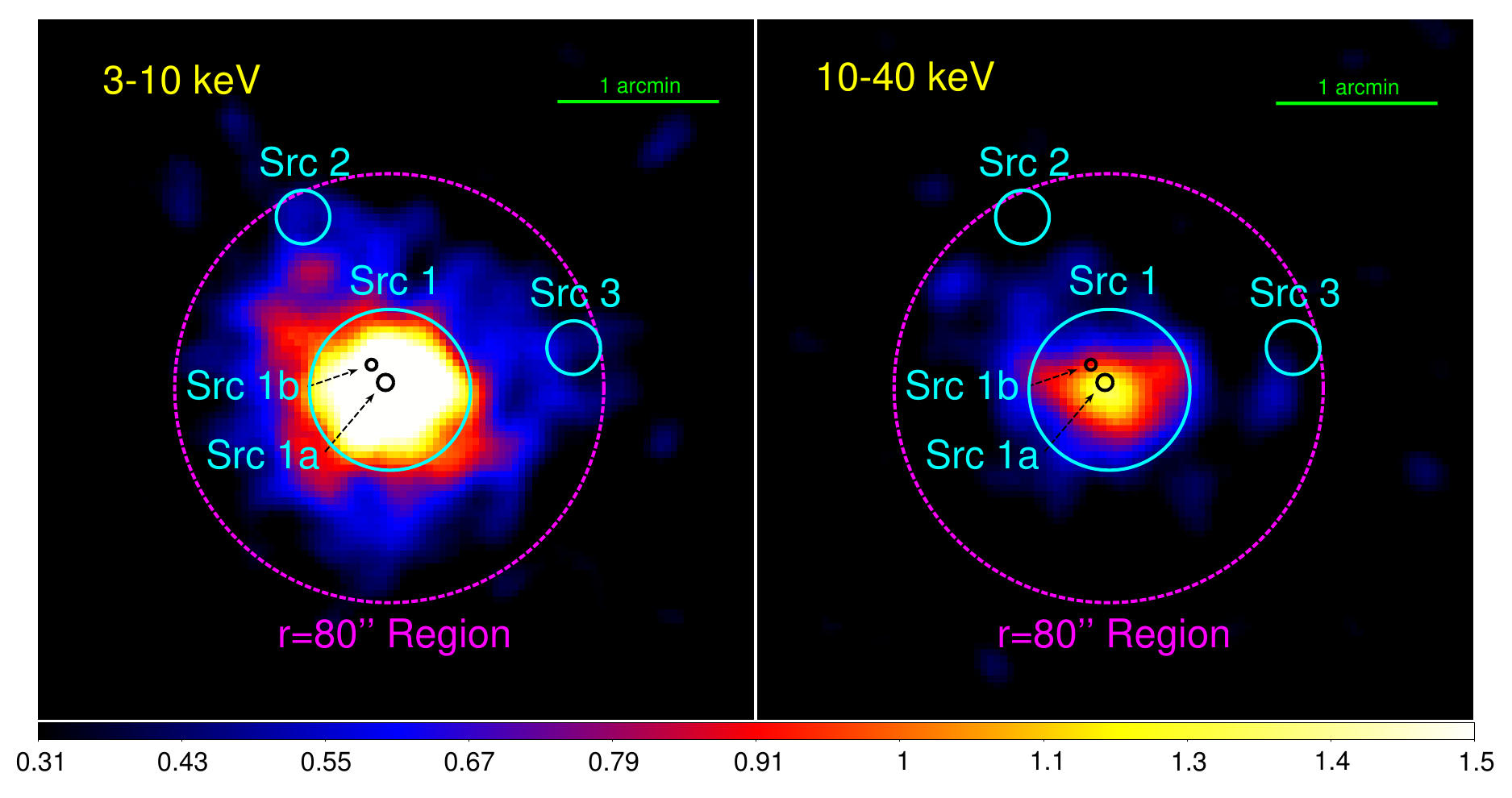}
\caption{The zoomed-in \nustar\ images of XMMU~J061804.3+222732 and the surroundings in 3--10~keV (left) and 10--40~keV (right). Both images are overlaid with source regions (cyan) for Src1 ($r=30\asec$) and Src 2 and 3 ($r=10\asec$), plus the $r=80\asec$ region covering all the three sources and other emission features. Source 2 and 3 are not significantly detected by \nustar, resulting in 1-$\sigma$ detection for source 2 and 2-$\sigma$ detection for source 3.
}
\end{figure*}

\section{Broadband X-ray Spectrum}

\subsection{Src 1 Spectrum}

We extracted both the \nustar\ and the \xmm\ spectra from a 30\asec\ radius region centered on Src 1 (the 30\asec\ cyan circle in Figure 3).
Background spectra were selected from local surrounding regions on the same detector chip.
We performed a joint spectral fit with \nustar\ and \xmm\ data using {\tt XSPEC} version 12.9.0 (Arnaud 1996).

To examine the SNR ejecta fragment model, we fit the 0.5--79~keV spectra with a model composed of a power-law and an optically thin thermal plasma component {\tt MEKAL}, both subject to foreground absorption, resulting in  {\tt constant*TBabs*(MEKAL+powerlaw)}.
It results in a satisfactory fit with $\chi^2_{\rm \nu}$=0.9 for $dof=328$ (left panel of Figure 4).
All the model parameters for {\tt TBabs}, {\tt MEKAL} abd {\tt powerlaw} are tied among the spectrum datasets, while {\tt constant} is frozen at 1 for the \nustar\ FPMA spectrum and left as a free parameter for all the other spectra.
The photon index of the power-law is tightly constrained as $\Gamma=1.72\pm0.08$; the temperature of the thin plasma model is $kT=0.19\pm0.04$~keV; the best-fit absorption column density is $N_{\rm H}=\rm (1.2\pm0.4)\times10^{22}~cm^{-2}$.
With this broad-band spectrum, we derived a softer spectrum for Src 1 than previous results of $\Gamma=1.48^{+0.20}_{-0.08}$ (Table 5 in \citealt{BB2003}) with improved error bars. 
The unabsorbed source flux is $F_{2-10\rm~keV}=\rm (6.1\pm 0.1)  \times 10^{-13} \rm~erg~cm^{-2}~s^{-1}$ in 2--10~keV and $F_{10-10\rm~keV}=\rm (8.0\pm 0.4)  \times 10^{-13} \rm~erg~cm^{-2}~s^{-1}$ in 10--40~keV.
The correspondent luminosities in 2--10~keV and 10--40~keV are $L_{2-10\rm~keV}=(0.72\pm0.02) \times10^{32}\rm~erg~s^{-1}$, and $L_{10-40\rm~keV}=(1.4\pm0.1)\times10^{32}\rm~erg~s^{-1}$, assuming a distance of 1.5~kpc.

To examine the PWN scenario, we also fit the spectra with blackbody plus a cutoff power-law ({\tt constant*TBabs*(bbody+cutoffpl)}), which models the combination of thermal emission from the neutron star surface, and non-thermal emission from the PWN.
This model results in an equally good fit with $\chi^2_{\rm \nu}$=0.8 for $dof=331$ (right panel of Figure 4).
The best-fit absorption column density is $N_{\rm H}=(0.9\pm0.2)\times10^{22}\rm~cm^{-2}$.
The blackbody temperature is derived to be $kT=0.11\pm0.01$~keV. 
The cutoff power-law has a photon index of $\Gamma=1.4\pm0.1$, consistent with previous measurements of $\Gamma=1.48^{+0.20}_{-0.08}$ using 1-10~keV data only, but with a cutoff energy at $E_{cut}=18^{+10}_{-5}$~keV.
Since we only one data bin above 20~keV with $S/N>3\sigma$, the current data cannot unambiguously distinguish between a power-law and a cutoff power-law.
However, we determined that the cutoff energy cannot be lower than 10 keV at 90\% confidence level.
The spectral fitting results for both the SNR ejecta and the PWN model are shown in Table 1.

Previous study has pointed out that Src 1 shows no flux variation from 2000 to 2006 \citep{Bykov2008}.
By comparing the 2--10~keV source flux in the 2015 \nustar\ observation with previous measurements by \chandra\ and \xmm, we conclude that Src 1 has maintained the same flux level for fifteen years from 2000 to 2015.

We also extracted \nustar\ and \xmm\ spectra from a larger region with $r=80\asec$ (the magenta circle in Figure 2), which contains all the three sources (Src 1-3) and the extended emission features in between the three sources.
The spectrum from the $r=80\asec$ circle is softer compared to the $r=30\asec$ circle centered on Src 1, resulting in $\Gamma=1.95\pm0.05$.
The flux of the $r=30\asec$ region makes up about 33\% of the total flux of the $r=80\asec$ region in 2--10~keV and $\sim50$\% above 10~keV.
This is consistent with previous results showing that there are thermal features in between sources 1-3, making the spectrum from a larger region softer. 

\begin{figure*}
\centering
\label{fig:src1spec}
\begin{tabular}{cc}
\includegraphics[angle=270, width=0.45\linewidth]{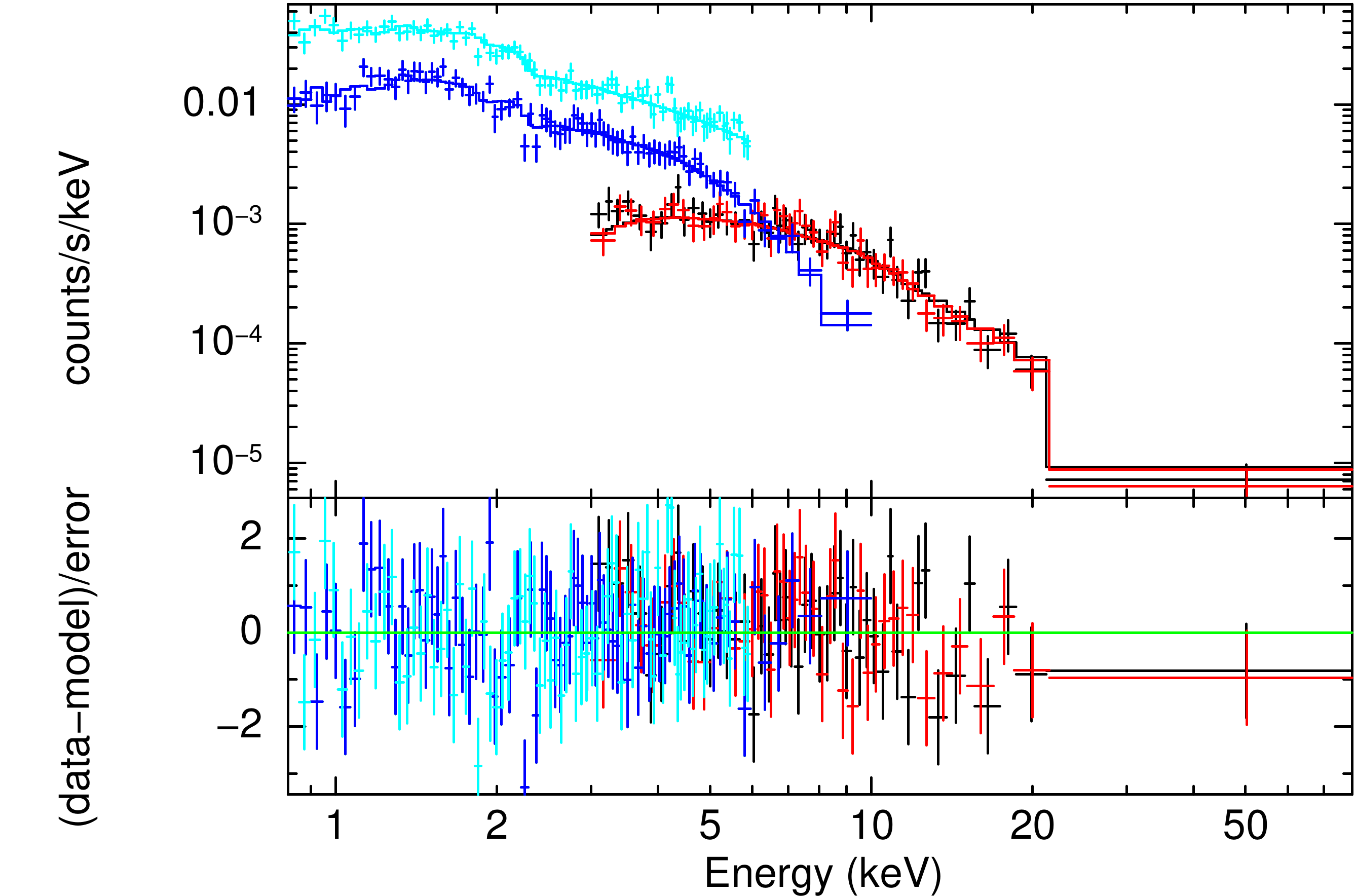} &
\includegraphics[angle=270, width=0.45\linewidth]{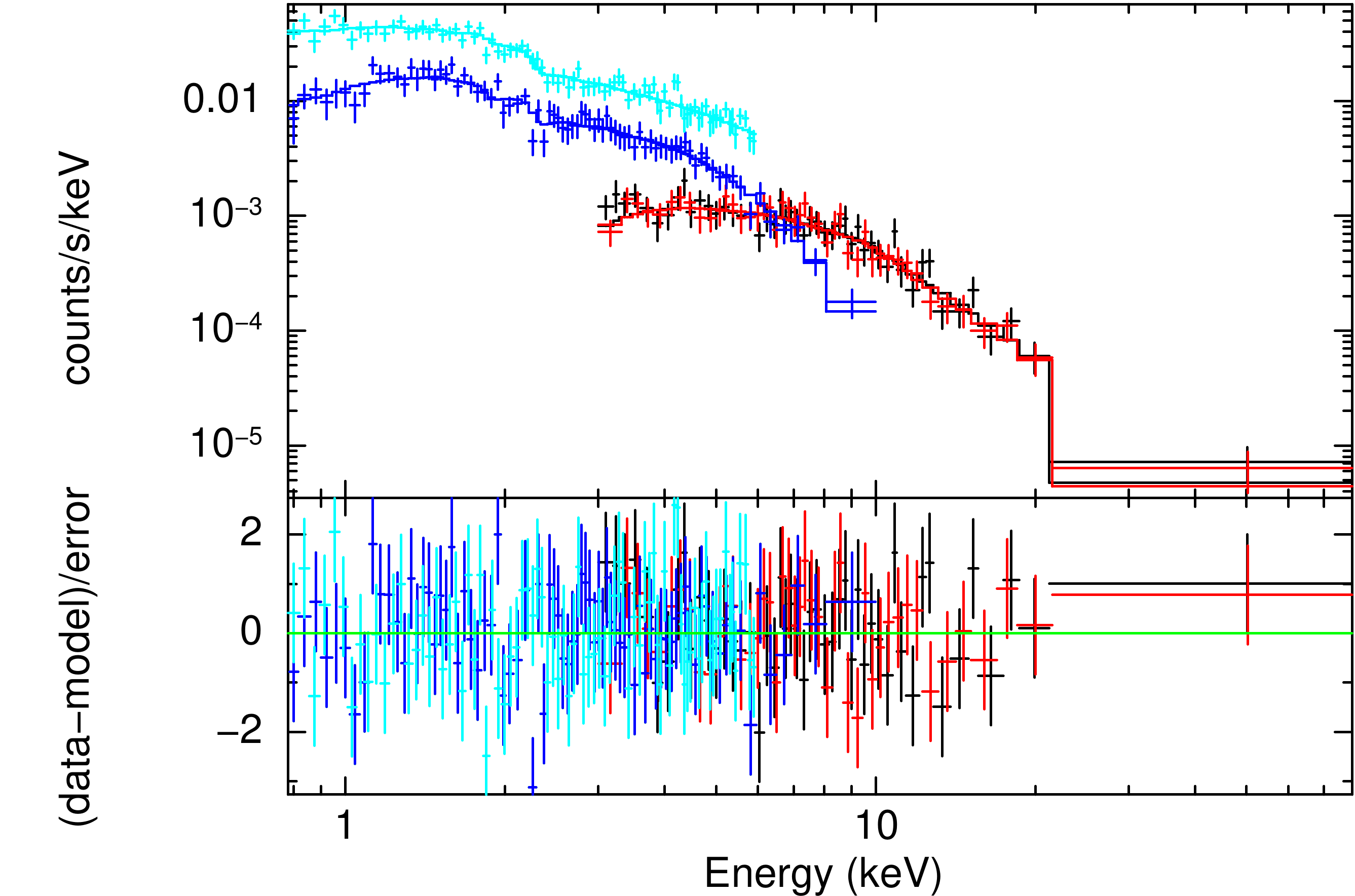}
\end{tabular}
\caption{
Joint \nustar\  (black for FPMA and red for FPMB) and \xmm\ (blue for combined MOS1 and MOS2, cyan for PN) spectra of Src 1 fitted to the PL+MEKAL model {\tt constant*tbabs*(MEKAL+powerlaw)} (left panel) and the BB+cutoffPL model {\tt constant*tbabs*(bbody+cutoffpl)} (right panel). The data points are with 1-$\sigma$ error bars, and the solid lines represent the best-fit model. The PL+MEKAL model results in a PL photon index of $\Gamma=1.72\pm0.08$ and $kT=0.19\pm0.04$~keV for the MEKAL component. The BB+cutoffPL model gives $\Gamma=1.48^{+0.20}_{-0.08}$ with cut-off energy at $E_{cut}=18^{+10}_{-5}$~keV and $kT=0.11\pm0.01$~keV for the blackbody component. Both models can fit to the broadband spectra very well. Due to the limited statistics above 20~keV, the current data cannot unambiguously distinguish between a power-law and a cutoff power-law. However, the cutoff energy cannot be lower than 10 keV at $>$90\% confidence level.
}
\end{figure*}

\begin{deluxetable*}{lccccc}[H]                                                                                                       
\tablecaption{Spectral fitting results for Src 1 with joint {\it NuSTAR} and {\it XMM} data.}
\tablecolumns{3}                                                                                                                    
\tablehead{ \colhead{Parameter}      & \colhead{Unit}                                       &  \colhead{MEKAL+power-law~$\rm Model^{a}$}   & \colhead{bbody+cutoffpl~$\rm Model^{b}$}  }
\startdata
$N_{\rm H}$                                      & 10$^{22}$~cm$^{-2}$                           & $1.2 \pm 0.4$                                                         & $0.9 \pm 0.2$                                        \\
& & & \\
$\Gamma$                                        &  \nodata                                                & $1.72\pm0.08$                                                       & $1.4\pm0.1$                                          \\
$E_{\rm cut}$                                    & keV                                                       & \nodata                                                                   & $18^{+10}_{-5}$                                     \\
$norm_{\rm pl/cutoffpl}$                   & ph~keV$^{-1}$~cm$^{-2}$~s$^{-1}$     & $(1.6\pm0.3)\times10^{-4}$                                    & $(1.2\pm0.1)\times10^{-4}$                   \\
& & & \\
$kT$                                                 &  keV                                                       & $0.19^{+0.04}_{-0.03}$                                          & $0.11\pm0.01$                                      \\
$norm_{\rm MEKAL/BB}$                & see note                                                & $3^{+7}_{-2}\times10^{-3}$                                    & $1.7^{+1.9}_{-0.7}\times10^{-5}$           \\
& & & \\
$F_{\rm 2-10keV}$                         & $10^{-13}$~egs~cm$^{-2}$~s$^{-1}$    & $6.1\pm0.1$                                                  & $6.1\pm0.1$  \\
$F_{\rm 10-40keV}$                       &  $10^{-13}$~egs~cm$^{-2}$~s$^{-1}$   & $8.0\pm0.4$                                                 & $5.6\pm0.4$  \\
& & & \\
$\chi^2_{\rm \nu}$ (dof)                     &                                          &  0.9 (328)                                                                              & 0.8 (331)                                                \\
\enddata
\tablecomments{
The goodness of fit is evaluated by $\chi^2_{\rm \nu}$ and the number of degrees of freedom (dof) given in parentheses. 
The errors are 90\% confidence. Unabsorbed Flux in 2--10~keV ($F_{\rm 2-10keV}$) and 10--79~keV ($F_{\rm 10-79keV}$) predicted by each model are reported here.\\
$^{a}$~MEKAL+power-law model: {\tt tbabs*(MEKAL+power-law)}. \\
$^{\ }$~The {\tt MEKAL} model is characterized by plasma temperature $kT$ in keV, normalization (norm) of\\
$^{\ }$~{\tt MEKAL} defined as $norm=10^{-14} [4\pi D_{A}(1+z)^{2}]^{-1}\int n_{e}n_{H}dV$,  where $D_{A}$ is the  angular diameter\\
$^{\ }$~distance to the source (cm), $n_{e}$ and $n_{H}$ are electron and H densities. \\
$^{b}$~bbody+cutoffpl model: {\tt tbabs*(blackbody+cutoffpl)}. \\
$^{\ }$~The {\tt bbody} model is characterized by temperature $kT$ in keV, and model norm defined as\\
$^{\ }$~$norm=$ $L_{39}^{2}/D_{10}^{2}$, where $L_{39}$ is  the source luminosity in units of $10^{39}$~erg~s$^{-1}$, and $D_{10}$ is\\
$^{\ }$~distance units of 10 kpc. The {\tt cutoffpl} is characterized by photon index $\Gamma$, e-folding energy of\\
$^{\ }$~exponential rolloff $E_{cut}$ (in keV), and the model norm in ph~keV$^{-1}$~cm$^{-2}$~s$^{-1}$ at 1~keV.\\
}
\label{specfit}
\end{deluxetable*}

\subsection{Src 2 and Src 3 Spectra and Long-term Variabilities}

Both Src 2 and Src 3 are not detectable in the 2015 \nustar\ observations, resulting in detection significance of 1-$\sigma$ and 2-$\sigma$, respectively.  
The 3-$\sigma$ detection limit for Src 2 and Src 3 is $F_{2-10\rm~keV} \sim 2\times10^{-14}$~erg~cm$^{-2}$~s$^{-1}$, which can be taken as the upper limit flux of these two sources.
We reanalyzed the 2004 \chandra\ observation to obtain spectral parameters for Src 2 and Src 3.
The regions used for source and background spectra extraction are showed in Figure 2.

The spectrum of Src 2 does not show any line feature.
We thus first fit its spectrum with a simple absorbed power-law model {\tt TBabs*powerlaw}. 
The best fit gives an absorption column density of $N_{\rm H}=\rm (1.0\pm0.2)\times10^{22}~cm^{-2}$ and a photon index of $\Gamma=2.4^{+0.3}_{-0.2}$ with $\chi^{2}$/dof=1.1/58.
We also applied an absorbed thermal plasma model {\tt TBabs*MEKAL} to the Src 2 spectrum, which fits equally well.
We fixed the abundance to solar, as it cannot be well constrained by the data.
The thermal plasma model gives the best-fit temperature of $kT=2.8^{+0.6}_{-0.5}$~keV.
These results are consistent with the \citet{Bykov2008} analysis.
However, we want to point out that the thermal plasma model cannot give an acceptable fit with the red shift fit to $z=0$, even after the abundance parameter set to free.
The thermal model instead requires a redshift of $z=0.09\pm0.01$.

Src 2 has shown time variability since the year of 2000.
Its 2--10~keV flux increased by a factor of $\sim4$ within five months in 2000, from $F_{2000,4}=0.5^{+0.4}_{-0.2}\times10^{-13}$~erg~cm$^{-2}$~s$^{-1}$ in April to $F_{2000,9}=(1.8\pm0.2)\times10^{-13}$~erg~cm$^{-2}$~s$^{-1}$ in September.
Then its flux decreased to $F_{2004}=(1.2\pm0.1)\times10^{-13}$~erg~cm$^{-2}$~s$^{-1}$ in 2004, and further decayed to $F_{2006}=(0.6\pm0.1)\times10^{-13}$~erg~cm$^{-2}$~s$^{-1}$ in 2006 \citep{Bykov2008}. 
Its 2015 flux in 2--10~keV is constrained to $F_{2015}\le0.2\times10^{-13}$~erg~cm$^{-2}$~s$^{-1}$ (3-$\sigma$ detection limit) by the \nustar\ observation, which suggests a flux decrease for an order of magnitude from 2000 to 2015.
There is no evidence for significant spectral variation during this time range. 

Similarly, we fit the Src 3 spectrum with both the power-law model and the thermal plasma model.
The power-law model gives a marginally acceptable fit to Src 3 spectrum with residuals between 3--4~keV ($\chi^{2}/dof=1.3/23$). 
The fit can be improved by adding an emission line, resulting in {\tt TBabs*(powerlaw+Gaussian)} ($\chi^{2}/dof=1.1/22$).
The best-fit gives a photon index of $\Gamma=1.8^{+0.4}_{-0.3}$, an emission line centroid energy of $E=3.7\pm0.1$~keV, and a slightly lower absorption of $N_{\rm H}=\rm (0.7^{+0.3}_{-0.2})\times10^{22}~cm^{-2}$ compared with Src 1 and Src 2.
The 2--10~keV flux of Src 2 in 2004 is $F_{2-10\rm~keV}=(1.2\pm0.1)\times10^{-13}$~erg~cm$^{-2}$~s$^{-1}$.
A thermal plasma model can also fit well to the Src 3 spectrum ($\chi^{2}/dof=1.1/22$), resulting in $N_{\rm H}=\rm (0.6^{+0.2}_{-0.1})\times10^{22}~cm^{-2}$ and $kT=10^{+10}_{-3}$~keV, but requiring a high redshift of $z=0.8\pm0.1$, which implies an extragalactic origin for Src 3.
Therefore, the line feature at $\sim3.7$~keV shall be taken with caution.
While a redshift of $z=0$ points to an Ar K line, a redshift of $z=0.8$ is consistent with a shifted Fe K line instead.

Src 3 also shows time variability, but with a different trend compared to Src 2.
The 2-10~keV flux of Src 3 increased from 2000 ($F_{2000}=(0.3\pm0.1)\times10^{-13}$~erg~cm$^{-2}$~s$^{-1}$) to 2004 ($F_{2004}=(0.6\pm0.1)\times10^{-13}$~erg~cm$^{-2}$~s$^{-1}$) and maintained at the same level until 2006 ($F_{2006}=(0.63\pm0.05)\times10^{-13}$~erg~cm$^{-2}$~s$^{-1}$).
However, this brightening trend does not seem to continue, as the Src 3 flux in 2015 cannot significantly exceed $F_{2015}=0.2\times10^{-13}$~erg~cm$^{-2}$~s$^{-1}$, comparable to its flux level back in 2000.

\subsection{Spectra of Two Point Sources in the FoV}

Here we also report the \nustar\ detection of two point sources in the FoV (obsID 30101058002).
The point source to the south of Src 1 is outlined with a $r=10\asec$ circle and listed as Ps 1 in Figure 1.
Ps 1 was detected by \xmm\ (named as XMMU~J061801.5+222233), and also by \chandra\ (catalogue source CXOGSG J061801.4+222228).
Within the \chandra/ACIS spatial uncertainty, there is an point-like optical counterpart GAIA DR1 id. 3377008814910613632.
This source is detected by \nustar\ at a significance level of 9-$\sigma$ with a total counts of 237 in 3--79~keV.
We performed a joint \chandra\ and \nustar\ spectral analysis for this source.
A simple absorbed power-law model can fit to the data very well, 
resulting in $N_{\rm H}=2.0^{+0.8}_{-0.7}\times10^{21}\rm~cm^{-2}$ and $\Gamma=2.0\pm0.2$ with $\chi_{\nu}^{2}=0.8$ for dof of 58.
The observed flux is $F_{2-10\rm~keV}=(1.3\pm0.6)\times10^{-13}\rm~erg~cm^{-2}~s^{-1}$ in 2--10~keV and $F_{10-40\rm~keV}=(1.4\pm0.8)\times10^{-13}\rm~erg~cm^{-2}~s^{-1}$ in 10-40~keV.
Alternatively, an absorbed bremsstrahlung model with $kT=5.8^{+1.6}_{-1.5}$~keV and $N_{\rm H}=0.8^{+0.5}_{-0.5}\times10^{21}\rm~cm^{-2}$ can fit to the data equally well.

The other point source to the east of Src 1 is denoted as Ps 2 in Figure 1.
This source is detected at 8-$\sigma$ significance level with a total counts of 173 in 3--79~keV.
Its spectrum is well fit by either a power-law with $\Gamma=1.9\pm0.5$, or a bremsstrahlung model with $kT=12^{+26}_{-6}$~keV.
The observed flux in 3--10~keV and 10--40~keV are $F_{3-10\rm~keV}=(8\pm4)\times10^{-14}\rm~erg~cm^{-2}~s^{-1}$ and $F_{10-40\rm~keV}=(1.2\pm0.7)\times10^{-13}\rm~erg~cm^{-2}~s^{-1}$.


\section{Discussion}

Within the SNR-MC interaction site of SNR IC~443, \nustar\ detected a hard X-ray source (Src 1) with a non-thermal spectrum ($\Gamma \sim 1.5-1.7$) and a stable X-ray luminosity of $L_{2-79\rm~keV}\sim3\times10^{32}$~erg~s$^{-1}$ from 2000 to 2015.
Two nearby X-ray sources (Src 2 and Src 3), both within 1.5\amin\ of Src 1, are found to demonstrate different time variabilities. 
Possible origins of isolated X-ray sources associated with SNR are PWNe, fast SN ejecta running in to molecular clouds and shocked molecular clumps.
Since Src 1-3 are located in the interaction site of SNR IC~443 and a surrounding cloud clump D, all the above scenarios are likely for their source nature.

Non-thermal X-ray emission from the SNR-MC  interaction site can also be produced by primary/secondary leptons.
However, we conclude that primary/secondary leptons cannot be a major source for the X-ray emission from Src 1. 
In the Appendix, we demonstrate our calculation on the contribution of the primary/secondary leptons to the X-ray emission of Src 1.
Therefore, in this section we focus on the three possible origins for isolated X-ray sources in the SNR-MC interaction site: 1) PWN 2) shocked molecular clumps; and 3) supernova ejecta fragments.

\subsection{PWN scenario}

There is already a known PWN (1SAX J0617.1+2221) in the IC~443 region, which has been assumed to be associated with IC~443.
However, since there is an overlapping SNR G189.6+3.3 \citep{AA1994}, the possibility of the existence of a second PWN in this complicated field cannot be ruled out.
The Src 1 X-ray spectrum of a blackbody with $kT \sim 0.1$~keV plus a power-law ($\Gamma \approx 1.7$) or a cut-off plower-law ($\Gamma \approx 1.4$ with $E_{cut} \sim18$~keV) is consistent with the PWN scenario. 
The blackbody emission can be explained by thermal emission from the surface of a neutron star.
The best-fit blackbody temperature $kT=0.11\pm0.01$~keV is consistent with those derived by \citet{Bykov2005, Bykov2008}, and is corresponding to about 3~km radius. 
The non-thermal power-law emission (w/wo energy cutoff) can arise from the nebula surrounding the central compact object, or from the magnetosphere of the neutron star.

We performed a pulsation search using the \nustar\ events collected within a $r=50\asec$ circle in different energy bands for Src 1.
A pulsation signal is not detected at 99\% confidence level.

\subsection{Shocked Molecular Clump}

Since the PWN scenario cannot be confirmed due to non-detection of pulsation signal, we next explored the shocked molecular clump scenario proposed by \citet{Bykov2000}.
{\cite{Bykov2000} investigated the non-thermal emission from a radiative SNR interacting with molecular clouds, which is similar to the case of IC 443. 
In the SNR-MC interaction site, the remnant drives both a forward shock into the dense molecular clump and a reverse shock into the radiative shell \citep{Chevalier99}. 
When the shock velocity is high enough, the hot shocked gas behind the shock can produce thermal X-ray emission with ISM/cloud abundance. This might explain the thermal plasma component from Src 1.

Non-thermal particles are accelerated at the MHD collisionless shock and then diffuse into the shocked clump and shell materials. 
These energetic particles can interact with the shocked materials and produce prominent non-thermal bremsstrahlung emission, producing sources bright in X-rays to MeV $\gamma$-rays in the SNR-MC interaction region. 
As long as the molecular clump density is much higher than that in the radiative shell, the emission from the interaction region is expected to be dominated by the shocked clump component. 
The spectrum of non-thermal electrons in a shocked clump is mainly shaped by the 1st-order Fermi acceleration. 
If large-scale turbulence driven by thermal and dynamical instability exists at the radiative shock, the 2nd-order Fermi acceleration can also affect the resulting electron spectrum. 
At low energies $\lesssim 100$~keV, the electrons suffer strong Coulomb loss, which is able to modify the electron spectrum. 
At high energies in the $\gamma$-ray window, the spectrum is expected to have a exponential-like cutoff. 
The maximum energy of the particles should be determined by either ion-neutral damping \citep{Drury96,PZ03} or limited particle acceleration time. 

A self-consistent model for the non-thermal emission from a shocked clump requires a proper treatment for the shock interaction between a remnant and molecular clouds \citep{Chevalier99}, plus a good recipe for particle injection and acceleration in an MHD collisionless shock \citep[][and reference therein]{Caprioli10}, which is beyond the scope of this paper. 
For simplification, we instead assume that the non-thermal electrons follow a power-law spectrum in momentum based on 1st-order Fermi acceleration, resulting in $N(E)dE\propto (E^2+2Em_ec^2)^{-\frac{s+1}{2}}\times(E+m_ec^2)dE$, where $E$ is electron kinetic energy and $s$ is electron spectral index. 
The normalization can be determined from the energy density of electrons with energies above 10 keV, $w_{e}(>10\ {\rm keV})$. 
Here we assume that the CR pressure is 10\% of the ram pressure, and that the electron to proton number ratio is $K_{ep}=0.01$ at momentum $p = 1\ {\rm GeV\ c}^{-1}$, and $w_{e}(>10\ {\rm keV})\approx 7.0\times10^{-10}n_{3}u^2_{s,2}\ {\rm erg\ cm}^{-3}$, where $n_3$ is the number density of clumps in $10^3\ {\rm cm}^{-3}$ and $u_{s,2}$ is the shock velocity in $100\ {\rm km\ s^{-1}}$.
At low energies, the power-law spectrum is modified by substantial Coulomb loss \citep{Vink08}. 
The 2nd-order Fermi acceleration is neglected here, which can overcome the Coulomb loss at low energies. 
As a result, we leave the Coulomb break as a free parameter in the data fitting.
We calculated the bremsstrahlung spectrum using analytic cross sections given in \citet{Haug1997}, with $e-e$ bremsstrahlung in the X-ray band neglected. 
With the above settings, the \nustar\ data can be fit well with a non-thermal bremsstrahlung with $s=2.0$ without invoking Coulomb loss (see blue dashed line in Figure~5), giving a shocked clump mass of $M_c\approx6.5M_{\odot}$.
In the case of $s=2.2$, we found that the data requires Coulomb loss with a break at $E_{\rm Cou}\approx15$~keV (black solid line in Figure~5), giving a molecular clump mass of $M_c\approx5.0M_{\odot}$.
The treatment here for the bremsstrahlung emission from a shocked clump is similar as that discussed in \citet{Uchiyama02}, and also agrees with the simulation results in \cite{Bykov2000} with more complicated
setup.

\begin{figure}
\label{fig:SED}
\includegraphics[angle=0, width=1.0\linewidth]{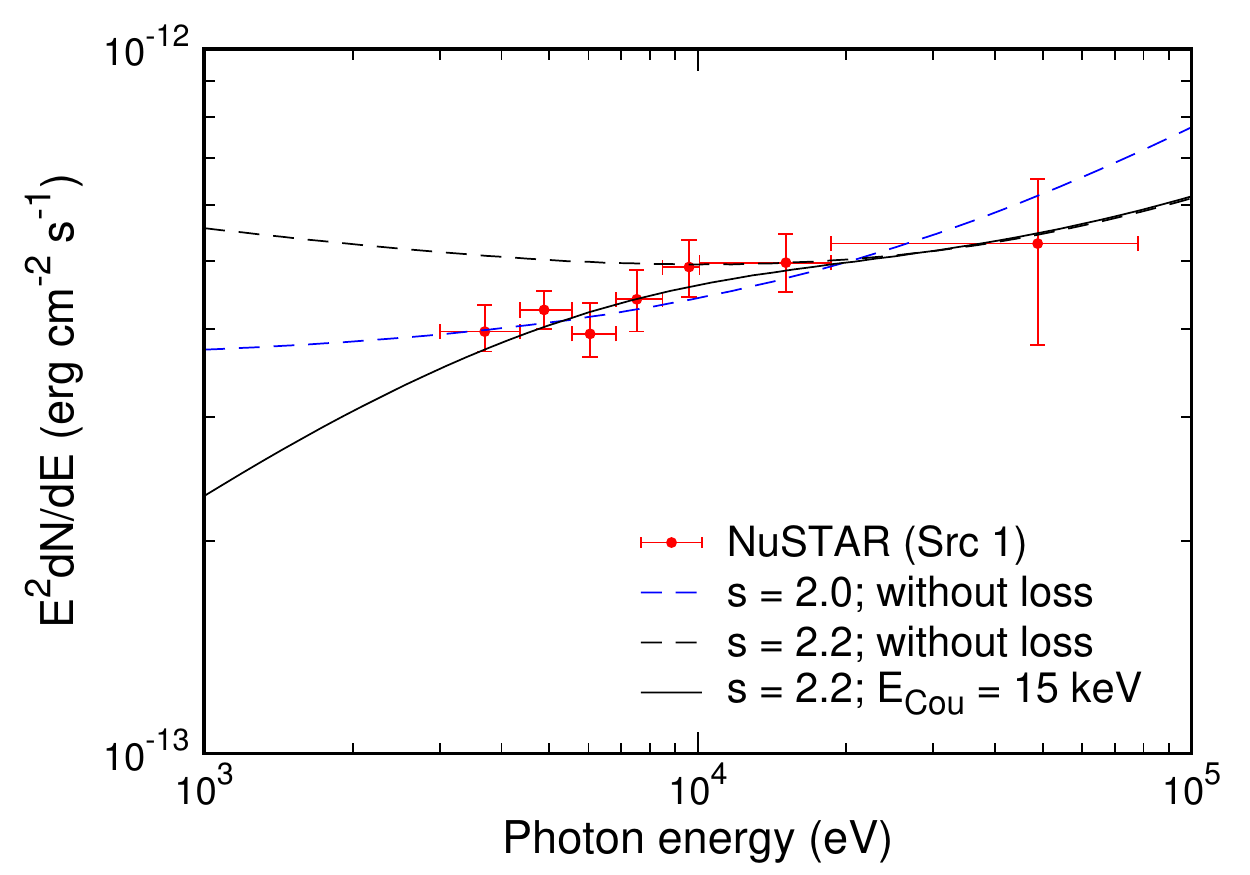}
\caption{The \nustar\ X-ray spectrum (noted by red dots with error bars) fit with shocked molecular clump model. A non-thermal bremsstrahlung emission with $s=2.0$ (blue dashed line), or a bremsstrahlung with $s=2.2$ with energy loss at $E_{Cou}=15$~keV (black solid line) are consistent with the \nustar\ data. $s=2.2$ without energy loss (black dashed line) can be ruled out.  
}
\end{figure}

\subsection{SNR Ejecta Fragments}

Fast moving X-ray knots were firstly discovered in Vela SNR with ROSAT by \cite{Aschenbach95}. 
Head-tail structure of apparent size of $8'.4\times 4'.1$ ($0.6\times 0.3$~pc at a distance of 250~pc) was revealed with high-resolution \chandra\ observation \citep{Miyata01}. 
Because of the relative high abundance in supernova nucleosynthesis products, the observed fast moving knots are interpreted as isolated supernova ejecta fragments, which are interacting with surrounding media and composed of heavy elements.

X-ray line and continuum emission from supernova ejecta fragments interacting with dense molecular clumps have been discussed for the situation in Vela SNR \citep{Bykov2002} and IC 443 \citep{Bykov2008}. 
It is believed that non-thermal particles are accelerated at the MHD collisonless shock and then diffuse into the shocked materials. 
The X-ray continuum emission of bremsstrahlung nature is produced in a similar way as in shocked molecular clumps (see Section 5.2). 
The X-ray line emission, a signature of SNR ejecta, is mainly due to K-shell ionization induced by energetic particles in the cold metallic part of ejecta fragments.
The resulting X-ray emission is most prominent when the ejecta fragments are colliding with a dense molecular clump. 
Therefore, SNR ejecta fragments and shocked molecular clumps are likely to exist next to each other.

Sub-structures of Src 1 revealed by \chandra\ show a few bright clumps (Src 1a, Src 1b etc, see Figure 2) embedded in an extended diffuse emission.
The Src 1a \chandra\ spectrum can be best fit with a power-law with $\Gamma=1.5^{+0.5}_{-0.4}$ plus a thermal plasma model with $kT=0.2^{+0.6}_{-0.1}$~keV \citep{Bykov2008}. 
The spectrum of the fainter sub-clump Src 1b can be described by a power-law with $\Gamma \sim 2$ with large error bars. 
The observed IR/X-ray morphology can possibly be explained by the interaction between IC 443 and clumpy molecular clouds, which involves multiple shocks with different velocities \citep{Bykov2008}.
The shocks are likely to be driven by the fast moving ejecta fragments, since a region next to Src 1a and 1b shows evidence for possible Si K-shell line. 
Considering the complicated morphology of Src 1, it is likely that it is composed of both SNR ejecta (or a PWN) and shocked molecular clumps.

\subsection{Nature of Src 2 and Src 3}

The other two isolated X-ray sources in the same SNR-cloud interaction site, Src 2  and Src 3, are not detectable in the 2015 \nustar\ data, resulting in 1-$\sigma$ and 2-$\sigma$ detection, respectively.
Src 2 showed a brightening in X-rays by a factor of four within five months in 2004, and then it X-ray flux keeps decreasing from 2004 to 2015.
The 2015 X-ray flux of Src 2 is $\sim10$\% of its highest X-ray flux in record. 
The 2--10~keV X-ray spectrum of Src 2 is featureless and can be equally fit with either a power-law with $\Gamma \sim2.4$, or a thermal plasma with $kT \sim 2.8$~keV and $z \sim 0.1$.
Its time variability and X-ray spectral shape are consistent with either a SNR ejecta fragment, or a shocked molecular clump illuminated by a SNR shock front in 2004. 
Alternatively, Src 2 could be a background galaxy.   

Src 3 demonstrates different X-ray characteristics in both time variability and spectrum.
The X-ray flux of Src 3 increased from 2000 to 2006, but dropped back to its previous X-ray flux in 2015. 
Its X-ray spectrum in 2--10~keV is well fit with a power-law with $\Gamma=1.8^{+0.4}_{-0.3}$ plus an emission line at $E=3.7\pm0.1$~keV.
Based on the detection of the line emission, Src 3 can be best interpreted as a SNR ejecta fragment if it is associated with IC 443.
Its X-ray spectrum can also be fit with a thermal plasma model with $kT=10^{+10}_{-3}$~keV with a redshift of $z=0.8$, in which case the 3.7~keV line shall be a shifted Fe K line. 
Therefore, an extragalactic origin, such as an AGN, is an alternative scenario for Src 3. 
Future \chandra\ monitoring of long-term variabilities of Src 2 and Src 3 will help to clarify their source nature. 

\subsection{Enlightenment on Nature of Galactic Center \\ Hard X-ray Sources}
The SNR ejecta / shocked clump interpretation of hard X-ray sources within SNR-MC interaction region can potentially explain a large population of hard X-ray sources in dense environments of star-forming regions, such as the Galactic center.
In the \nustar\ Galactic center survey, we detected a few hard X-ray sources in the SNR-MC interaction sites \citep{Zhang2014, Nynka2015}. 
Among them, G359.97-0.038 is most likely arising from SNR-MC interaction.

G359.97-0.038 is located at the interaction site of the Galactic center SNR Sgr A East and a giant molecular cloud M-0.02-0.07 (a.k.a. the 50 km~s$^{-1}$ cloud), strongly indicated by the presence of OH 1720~MHz maser \citep{PS2006}.
It shows a similar hard continuum spectrum ($\Gamma=1.3\pm0.3$) as Src 1 in the IC 443 system, with no significant line features. 
Therefore, as discussed in Sections 5.2 and 5.3, G359.97-0.038 can be a shocked molecular clump.
While \nustar\ detected a few X-ray sources in the vicinity of Sgr A East, \chandra\ revealed many more isolated X-ray sources in the interaction site of the Sgr A East and M-0.02-0.07 cloud system.
In future works we will apply the SNR ejecta/shocked clump models developed here on the unidentified X-ray sources around Sgr A East.
Considering the rich environment for star forming regions at the Galactic center, a large population of sources with hard X-ray spectra could bear the same source nature.

\acknowledgements
This work is supported under NASA Contract No. 6926645. This work is based on observations made with \nustar\ mission, a project led by the California Institute of Technology, managed by the Jet Propulsion Laboratory, and funded by NASA. We thank the \nustar\ Operations, Software and Calibration teams for support with the execution and analysis of these observations. This research has made use of the \nustar\ Data Analysis Software (NuSTARDAS) jointly developed by the ASI Science Data Center (ASDC, Italy) and the California Institute of Technology (USA). This research has also made use of data obtained with \xmm, an ESA science mission with instruments and contribution directly funded by ESA Member States and NASA, and also \chandra, a NASA science mission directly funded by NASA. S.Z. acknowledges support from NASA through the Smithsonian Astrophysical Observatory contract to MIT. D.P. acknowledges support from NASA through an Einstein fellowship (PF6-170156). Z.Y.Z. acknowledges support from the European Research Council in the form of the Advanced Investigator Programme, 321302, COSMICISM.

\appendix
\subsection{Contribution from SNR-accelerated Primary and Secondary Leptons to Src 1 X-ray Emission}

SNRs are believed to be the most important cosmic-ray accelerator up to the knee energy of $\sim 10^{15}$eV. 
Cosmic-ray electrons and protons accelerated at the remnant shock can produce multi-wavelength non-thermal emission from radio to $\gamma$-ray band through synchrotron, bremsstrahlung, inverse Compton (IC) and pion-decay emission mechanisms.  
Recently $\gamma$-ray emission has been detected in IC~443 by AGILE \citep{Tavani10} and Fermi \citep{Abdo2010}. 
There is still debate about the origin of the observed $\gamma$-ray emission.  
One explanation involves direct interaction between the remnant and dense molecular clouds \citep{Uchiyama10, TC14,TC15}, while the other propose that observed emission is from nearby clouds illuminated by escaping comic-ray particles \citep{LC10,Ohira11}. 
Despite above debate, it is generally accepted that the $\gamma$-ray emission is produced by hadronic interaction between accelerated protons and nuclei in the dense molecular clump. 
Since the characteristic $\pi^{0}$-decay signature for proton proton interaction is claimed to be detected in IC 443 recently \citep{Ackermann2013}. 
Here we provide a SED fitting for the multi-wavelength emission from IC443 and compare it with Src 1 X-ray emission measured by \nustar. 
For simplification, we assume the accelerated protons and electrons follow a smooth broken power-law. 
We also assume that the electron to proton number ratio $k_{ep}=0.01$ at momentum $p=1$GeV/c.

The comic-ray leptons producing broadband non-thermal emission could come from either direct SNR shock acceleration (referred as primary leptons) or via hadronic processes (referred as secondary leptons). 
The shock-accelerated relativistic protons and irons may interact with nuclei in the dense molecular cloud ($p$-$p$ collision), which creates both neutral pions that decay into two $\gamma$-ray photons and also charged pions  that decay into secondary leptons and neutrinos.
At the same time, the secondary leptons can radiate broadband non-thermal emission through synchrotron and bremsstrahlung processes in the ambient gas.
Previous works have suggested that X-ray emission from SNR-cloud interaction systems could be explained by such mechanism.
We therefore first investigated whether the secondary electrons contribute to the \nustar\ X-ray sources within IC~443.  

Firstly, we determined the spectrum of the parent protons by fitting the $\gamma$-ray data.
We adopted the following broken power-law to model the $\gamma$-ray data, 
\begin{eqnarray}
f_p(t_{\rm age}, E_p)&=&\frac{dN_p(E_p)}{dE_p}=N_p E_{p}^{-p_1} {\rm exp}\left[  -\left(\frac{E}{E_{p,cut}}\right) \right] \nonumber \\
&&\times \left[ 1+\left(\frac{E_p}{E_{p,br}}\right)^{2}  \right]^{\frac{p_2-p_1}{2} } 
\end{eqnarray}
where $t_{\rm age}$ is the SNR age, $E_p$ is the proton energy; $E_{p,br}$ is the break energy; $p_1$ and $p_2$ are the proton indices before and after break, respectively; 
$E_{p,cut}$ is the cutoff energy and is fixed at $E_{p,cut}=3$~PeV; the normalization $N_p$ is determined by $W_{p}$, the total energy in protons with energies above 1~GeV.
We adopted the following parameters for IC~443 in this calculation: distance of $d=1.5$~kpc, average gas density of $n_{\rm H}=200\ {\rm cm}^{-3}$ and age of $t_{age}=10^4$~yr. 
We assumed that the shock acceleration efficiency is a constant,
and ignored the energy loss due to the long lifetime of protons ($\tau_{pp}\sim 6\times10^7 (n_{\rm H}/1\ {\rm cm}^{-3})^{-1}$ yr). 
Thus, at a given time $t$ for $t\leq t_{\rm age}$, the proton distribution in the emission zone can be expressed as $f_p(t,E_p)=t\cdot f_p(t_{\rm age},E_p)/t_{\rm age}$.
The best-fit model is represented by the black solid line in Figure~\ref{fig:sed}.
The corresponding parent proton spectrum has a break energy of $E_{p,bre}=180$~GeV, proton spectral indices of $p_1=2.2$, $p_2=3.2$, and a total proton energy of $W_p=4\times10^{48}\ (n_{\rm H}/200\ \mathrm{cm}^{-3})^{-1}$~erg. 

The injection rate of secondary electrons $Q_e(t, E_e)$ can be given by
$Q_e(t, E_e) = c\,n_{\rm H}\int \sigma_{inel}(E_{p})f_p(t,E_p)dE_p$, where $c$ is the speed of light.
We calculated the spectrum of the injected secondary electrons $Q_e(t, E_e)$ using the parametrization presented in \citep{Kelner2006}. 
Due to energy loss of electrons, the accumulated electron distribution function at given time t, $f_e(t, E_e)$, can be obtained by solving the following equation
\begin{equation}\label{eq:dnde}
\frac{\partial f_e(t,E_e)}{\partial t} =\frac{\partial}{\partial E_e}\left[\left(\frac{dE_e}{dt}\right)f_e(t,E_e)\right]+Q_e(t,E_e)
\end{equation}

We then calculated the synchrotron and bremsstrahlung emission from secondary electrons assuming a magnetic field strength of $B=100\ \mu$G.
The resultant broadband non-thermal emission of secondary leptons from $p-p$ interaction are represented by dashed orange (synchrotron) and blue (bremsstrahlung) lines in Figure~\ref{fig:sed}. 
It is apparent that synchrotron and bremsstrahlung emission from secondary electrons cannot be the dominant physical process giving rise to the hard X-ray emission from Src 1.
While the synchrotron radiation from the secondary electrons strongly depends on the magnetic field strength $B$, 
the bremsstrahlung emissivity is sensitive to the ambient gas density $n_{\rm H}$. 
We therefore varied the values for both $B$ and $n_{H}$ by one order of magnitude, but it still fails to fit Src 1's spectrum.

We next considered the contributions of primary leptons (leptons directly accelerated by the SNR) as well.
The resultant broadband non-thermal emission of primary leptons are represented by solid orange (synchrotron) and blue(bremsstrahlung)  lines in Figure~\ref{fig:sed}. 
Our results show that the primary leptons are much more efficient in producing broadband emission from radio to $\gamma$-rays in the case of IC~443.
The synchrotron emission from the primary leptons could well fit to the radio emission from IC 443.
However, the combination of synchrotron and bremsstrahlung emission from primary electrons is at least one order of magnitude lower compared to the hard X-ray emission from Src 1.
Therefore, we conclude that both the leptons accelerated by the SNR and the secondary leptons from $p$-$p$ interaction in the cloud cannot be the major source for the bright hard X-ray emission in the IC 443 SNR-cloud interaction site.

\begin{figure}[H]
\centering
\includegraphics[height=55mm]{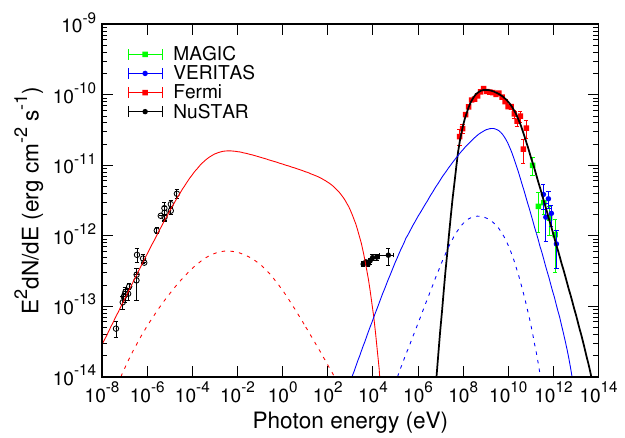}
\caption{Broadband energy spectrum of IC 443 with data obtained in radio by Clark Lake TPT telescope (black open circles, \citealt{Erickson1985}), X-rays by \nustar\ (black solid circles, this work), GeV $\gamma$-rays by \fermi (red squares, \citealt{Ackermann2013} and TeV $\gamma$-rays by MAGIC (green square, \citealt{IC443.MAGIC.2007}) and VERITAS (blue circles, \citealt{IC443.VERITAS.2009}). The black solid line shows the best-fit model for the emission rising from the $\pi_{0}$ decay of the parent protons after $p$-$p$ interaction. The dashed lines in orange and blue are best-fit models for the synchrotron and bremsstrahlung emission from secondary electrons, respectively. The solid orange and blue lines represent synchrotron and bremsstrahlung emission from the primary electrons. The sum of non-thermal emission from primary and secondary leptons is one order of magnitude lower than the bright X-ray emission from Src 1 in the \nustar\ band. Therefore, this mechanism cannot be a major source for Src 1 X-ray emission.} 
\label{fig:sed}
\end{figure}

\end{document}